\documentclass[final,5p,times,twocolumn]{elsarticle}

\usepackage{microtype}
\usepackage{amsmath,amssymb}
\usepackage{siunitx}
\usepackage{hyperref}
\usepackage{cleveref}

\hypersetup{colorlinks=true, linkcolor=blue, citecolor=blue, urlcolor=blue}

\journal{Proceedings of the 18th International Workshop on Top Quark Physics (TOP2025)}

\begin{document}

\begin{frontmatter}

\title{EFT Results in the Top Quark Sector in CMS}

\author[nd]{Andrea Piccinelli}
\address[nd]{University of Notre Dame, Notre Dame, Indiana, USA}

\begin{abstract}
The CMS programme of indirect searches in the top-quark sector interprets precision measurements in the Standard Model Effective Field Theory (SMEFT) framework. This proceedings summarises recent CMS results highlighted in the TOP2025 talk: a search for \(CP\) violation in \(t\bar t Z\) and \(tZq\) production using CP-odd observables constructed with physics-informed machine learning, a measurement that disentangles the flavour structure of electroweak SMEFT couplings in multilepton final states, and a Run~2 combination of complementary top+X measurements. We close with a brief outlook on the expected sensitivity gains at the high-luminosity LHC.
\end{abstract}


\end{frontmatter}

\section{Introduction}
Direct searches for new particles at the LHC have not yet revealed physics beyond the Standard Model (SM). If the new-physics scale \(\Lambda\) is above the energies directly accessible, its effects can be parameterised by an effective field theory expansion in higher-dimensional operators built from SM fields and symmetries. In the SMEFT, the effective Lagrangian is written as
\begin{equation}
\mathcal{L}_{\mathrm{eff}} = \mathcal{L}_{\mathrm{SM}} + \sum_{i} \frac{c_i}{\Lambda^2}\,\mathcal{O}_i + \cdots,
\end{equation}
where the Wilson coefficients (WCs) \(c_i\) encode the strength of dimension-6 operators \(\mathcal{O}_i\). The top quark is a key probe of such effects due to its large mass and strong couplings to the Higgs and electroweak sectors. Recent CMS measurements constrain SMEFT interactions using multi-dimensional likelihood fits to multichannel data sets, enabling both targeted studies (e.g. \(CP\)-odd effects) and more global interpretations.

\section{Search for \texorpdfstring{\(CP\)}{CP} violation in \texorpdfstring{\(t\bar t Z\)}{ttZ} and \texorpdfstring{\(tZq\)}{tZq}}
CMS performed a search for \(CP\) violation in events with top quarks produced in association with a \(Z\) boson in final states with at least three charged leptons and additional jets, targeting \(t\bar t Z\) and \(tZq\) production. The analysis uses pp collision data collected in 2016--2018 at \(\sqrt{s}=\SI{13}{\tera\electronvolt}\) and in 2022 at \(\sqrt{s}=\SI{13.6}{\tera\electronvolt}\), for a total integrated luminosity of \(\SI{173}{\femto\barn^{-1}}\) \cite{CMS_CPViolation_ttZtZq_2025}. Two \(CP\)-odd dimension-6 operators are considered, \(\mathcal{O}^{I}_{tW}\) and \(\mathcal{O}^{I}_{tZ}\), corresponding to the imaginary parts of electroweak dipole operators \cite{CMS_CPViolation_ttZtZq_2025}.

A distinguishing feature of this measurement is the use of observables that are odd under the \(CP\) transformation. In the SM, the distributions of these observables are predicted to be symmetric around zero, while \(CP\)-violating interactions would introduce asymmetries \cite{CMS_CPViolation_ttZtZq_2025}. Physics-informed machine learning is employed to construct the CP-odd observables in the partially reconstructed multilepton final state \cite{CMS_CPViolation_ttZtZq_2025}.

\begin{figure*}[t]
  \centering
  \includegraphics[width=0.98\textwidth]{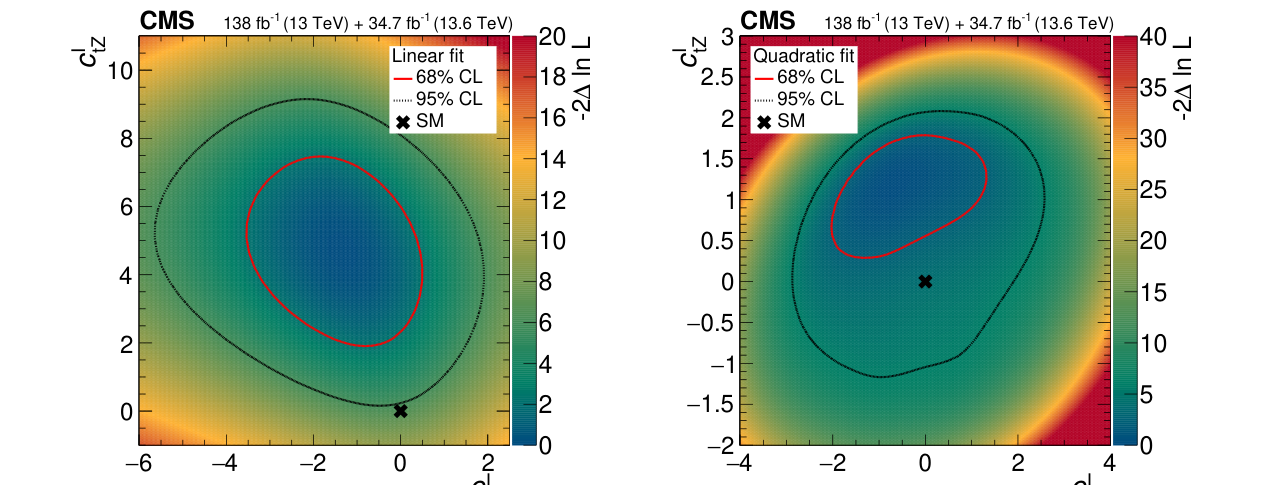}
  \caption{Likelihood scans in the \((c^{I}_{tW},\,c^{I}_{tZ})\) plane for the \(CP\)-violation search in \(t\bar t Z\) and \(tZq\) production. Left: linear EFT contributions only. Right: linear and quadratic contributions. The SM point is indicated by a cross. Adapted from \cite{CMS_CPViolation_ttZtZq_2025}.}
  \label{fig:cpv}
\end{figure*}

The results are consistent with the SM within two standard deviations. At 95\% confidence level, the limits on the associated WCs are
\( -2.7 < c^{I}_{tW} < 2.5\) and \( -0.2 < c^{I}_{tZ} < 2.0\) \cite{CMS_CPViolation_ttZtZq_2025}. A mild preference for positive \(c^{I}_{tZ}\) is observed in a linear-only interpretation, corresponding to a local significance of 2.5 standard deviations with respect to the SM hypothesis \cite{CMS_CPViolation_ttZtZq_2025}. \Cref{fig:cpv} shows the likelihood contours in the \((c^{I}_{tW},\,c^{I}_{tZ})\) plane for fits including only linear EFT contributions and including both linear and quadratic terms.

\section{Flavour structure of electroweak SMEFT couplings in multilepton final states}
Beyond constraining overall deviations from the SM, SMEFT interpretations can probe flavour-dependent patterns. CMS reported the first measurement that disentangles the flavour structure of selected dimension-6 operators affecting quark--\(Z\) couplings by simultaneously probing interactions with different quark generations \cite{CMS_FlavorStructure_EFT_Multilepton_2025}. The analysis targets multilepton final states with at least three electrons or muons, and performs a simultaneous fit in signal regions enriched in \(t\bar t Z\), \(WZ\), and \(ZZ\) production using the \(Z\)-boson transverse momentum distribution \cite{CMS_FlavorStructure_EFT_Multilepton_2025}. The data correspond to \(\SI{138}{\femto\barn^{-1}}\) collected at \(\sqrt{s}=\SI{13}{\tera\electronvolt}\) (2016--2018) \cite{CMS_FlavorStructure_EFT_Multilepton_2025}.

The probed operators modify electroweak vector couplings in the Warsaw basis, including \(\mathcal{O}_{\phi q}^{(1)}\), \(\mathcal{O}_{\phi q}^{(3)}\), \(\mathcal{O}_{\phi u}\), and \(\mathcal{O}_{\phi d}\), with separate parameters for light-quark generations (first and second) and the third generation \cite{CMS_FlavorStructure_EFT_Multilepton_2025}. By combining \(t\bar t Z\) with diboson channels, the analysis gains sensitivity to light-quark couplings that enter through initial-state radiation and dominant diboson production, while retaining sensitivity to top-quark couplings through diagrams where the \(Z\) is radiated from the top quark \cite{CMS_FlavorStructure_EFT_Multilepton_2025}.

\begin{figure}[t]
  \centering
  \includegraphics[width=\linewidth]{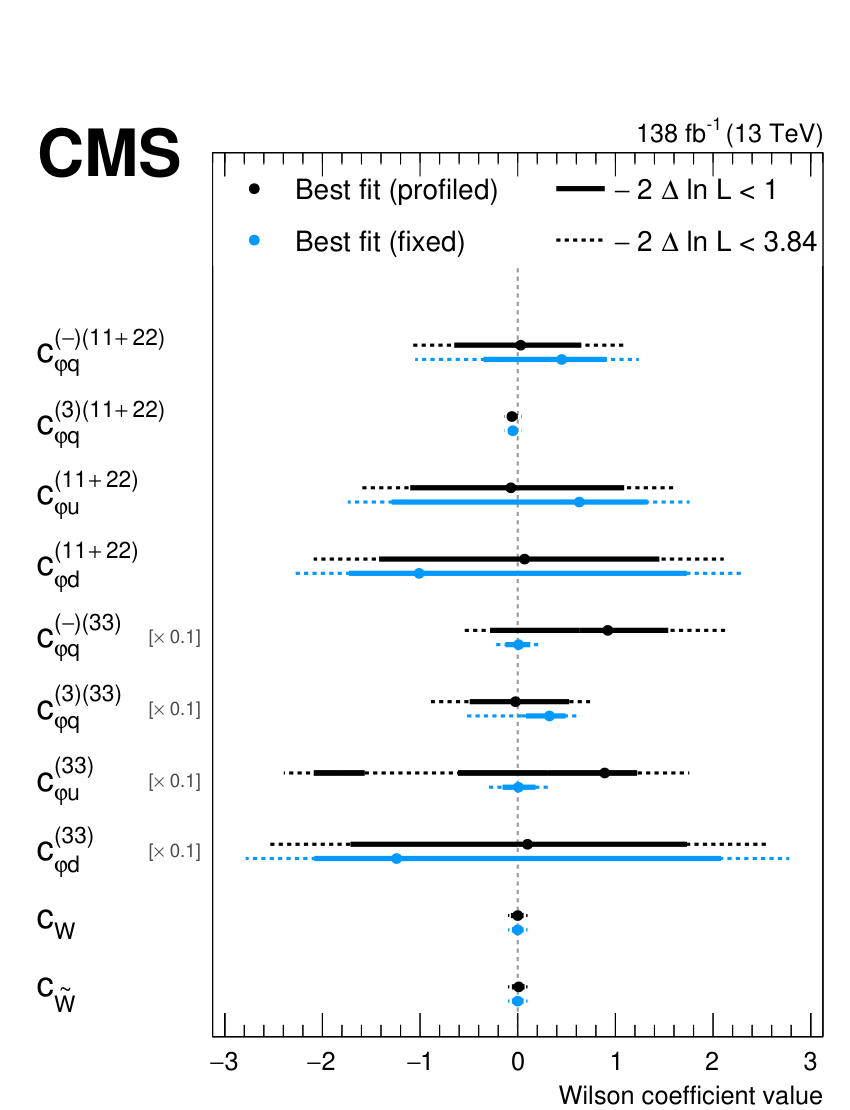}
  \caption{Summary of best-fit values and confidence intervals for selected WCs in the flavour-structure measurement in multilepton final states. Results are shown for scans with other coefficients fixed to zero and for profiled fits. Heavy-quark couplings are scaled by 0.1 for visibility. Adapted from \cite{CMS_FlavorStructure_EFT_Multilepton_2025}.}
  \label{fig:flavor}
\end{figure}

The observed data are consistent with the SM, and limits are set on the flavour-split WCs. \Cref{fig:flavor} summarises the best-fit values and confidence intervals for the selected parameters, comparing fits where other WCs are fixed to zero versus profiled in the likelihood \cite{CMS_FlavorStructure_EFT_Multilepton_2025}.

\section{Combinations and projections}
Combining complementary final states enhances sensitivity and improves the robustness of SMEFT interpretations. CMS presented a Run~2 statistical combination of two indirect searches: a ``boosted'' analysis targeting \(t\bar t\) production with a hadronically decaying boson (\(Z\) or \(H\)) at high transverse momentum, and a multilepton analysis targeting associated top-quark production with additional leptons \cite{CMS_PAS_TOP_24_004_2025}. Both inputs use \(\SI{138}{\femto\barn^{-1}}\) at \(\sqrt{s}=\SI{13}{\tera\electronvolt}\) and constrain eight WCs associated with independent dimension-6 operators in a simultaneous fit \cite{CMS_PAS_TOP_24_004_2025}. The combination improves the constraints by up to 10\% with respect to either analysis alone \cite{CMS_PAS_TOP_24_004_2025}.

Looking forward, CMS provided projections for indirect SMEFT sensitivity in top-quark production with additional leptons at the high-luminosity LHC, based on a global Run~2 EFT analysis extrapolated to \(\SI{6000}{\femto\barn^{-1}}\) for a combined ATLAS+CMS dataset \cite{CMS_NOTE_2025_008_Projections}. Under the considered uncertainty scenarios, the expected constraints on the WCs improve by roughly a factor of 2--4 compared to Run~2 \cite{CMS_NOTE_2025_008_Projections}. These studies motivate continued development of multi-process combinations, improved theoretical modelling, and higher-granularity phase-space exploration as the dataset grows.

\section{Conclusions}
CMS continues to advance an extensive SMEFT programme in the top-quark sector, spanning targeted symmetry tests and global multichannel interpretations. The recent search for \(CP\) violation in \(t\bar t Z\) and \(tZq\) production introduces CP-odd observables constructed with physics-informed machine learning and sets new limits on \(c^{I}_{tW}\) and \(c^{I}_{tZ}\) \cite{CMS_CPViolation_ttZtZq_2025}. Complementary multilepton measurements probe the flavour structure of electroweak SMEFT couplings by simultaneously constraining light- and heavy-quark interactions with the \(Z\) boson \cite{CMS_FlavorStructure_EFT_Multilepton_2025}. Run~2 combinations of top+X measurements already provide measurable gains in sensitivity \cite{CMS_PAS_TOP_24_004_2025}, and HL-LHC projections indicate substantially improved reach for indirect new-physics searches in the coming decade \cite{CMS_NOTE_2025_008_Projections}.

\section*{Acknowledgements}
I thank the CMS Collaboration and the TOP physics and EFT groups for their work on the results summarised here, and the TOP2025 organisers for the opportunity to present them \cite{Piccinelli_TOP2025_Talk}.

\bibliographystyle{elsarticle-num}
\bibliography{references}

\begin{thebibliography}{1}
\expandafter\ifx\csname url\endcsname\relax
  \def\url#1{\texttt{#1}}\fi
\expandafter\ifx\csname urlprefix\endcsname\relax\def\urlprefix{URL }\fi
\expandafter\ifx\csname href\endcsname\relax
  \def\href#1#2{#2} \def\path#1{#1}\fi

\bibitem{CMS_CPViolation_ttZtZq_2025}
{CMS Collaboration}, Search for {CP} violation in events with top quarks and
  {Z} bosons at $\sqrt{s}=13$ and $13.6\,\mathrm{TeV}$, Phys. Lett. BSubmitted;
  arXiv:2505.21206 (2025).
\newblock \href {http://arxiv.org/abs/2505.21206} {\path{arXiv:2505.21206}}.

\bibitem{CMS_FlavorStructure_EFT_Multilepton_2025}
{CMS Collaboration}, Probing the flavour structure of dimension-6 {EFT}
  operators in multilepton final states in proton-proton collisions at
  $\sqrt{s}=13\,\mathrm{TeV}$, JHEPSubmitted; arXiv:2507.17498 (2025).
\newblock \href {http://arxiv.org/abs/2507.17498} {\path{arXiv:2507.17498}}.

\bibitem{CMS_PAS_TOP_24_004_2025}
{CMS Collaboration}, Combination of exclusion limits on modified couplings
  between top quarks and heavy bosons in the effective field theory framework,
  CMS Physics Analysis Summary CMS-PAS-TOP-24-004, CERN, 31 March 2025 (2025).

\bibitem{CMS_NOTE_2025_008_Projections}
{CMS Collaboration}, Projections for indirect searches for new physics in the
  production of top quarks in association with additional leptons, CMS Note
  CMS-NOTE-2025/008, CERN, 17 June 2025 (v3 08 July 2025) (2025).

\bibitem{Piccinelli_TOP2025_Talk}
A.~Piccinelli, {EFT Results in the top quark sector in CMS} (talk at the 18th
  international workshop on top quark physics, top2025), Conference
  presentation (slides), hanyang University, Seoul (KR), September 2025 (2025).

\end{thebibliography}

\end{document}